\documentclass{llncs}

\usepackage[usenames]{color}

\usepackage{comment, graphicx, latexsym}
\usepackage{amsmath, amssymb, amsfonts, cite}

\begin{document}

\title{The Hidden Subgroup Problem and Post-quantum Group-based Cryptography}
\author{Kelsey Horan\inst{1}  \and Delaram Kahrobaei\inst{2}, \inst{3}}
\institute{ The Graduate Center, CUNY, USA \\ \email{khoran@gradcenter.cuny.edu} \and The Graduate Center and NYCCT, CUNY \and New York University, USA \\ \email{dk2572@nyu.edu} \thanks{Research of Delaram
Kahrobaei was partially supported by a PSC-CUNY grant from the CUNY
research foundation. Research
of Delaram Kahrobaei  was also supported by
the ONR (Office of Naval Research) grant N000141512164. We thank E. Kashefi and L. Perret for discussions and hospitality in summer 2017 at University of Sorbonne.}}

\maketitle

\begin{abstract}
In this paper we discuss the Hidden Subgroup Problem (HSP) in relation to post-quantum cryptography. We review the relationship between HSP and other computational problems discuss an optimal solution method, and review the known results about the quantum complexity of HSP. We also overview some platforms for group-based cryptosystems. Notably, efficient algorithms for solving HSP in the proposed infinite group platforms are not yet known.

\end{abstract}


\section{Introduction} \label{sec:intro}

In August 2015 the National Security Agency (NSA) announced plans to upgrade security standards; the goal is to replace all deployed cryptographic protocols with quantum secure protocols. This transition requires a new security standard to be accepted by the National Institute of Standards and Technology (NIST). Proposals for quantum secure cryptosystems and protocols have been submitted for the standardization process. There are six main primitives currently proposed to be quantum-safe: (1) lattice-based (2) code-based (3) isogeny-based (4) multivariate-based (5) hash-based, and (6) {\it group-based} cryptographic schemes.

One goal of cryptography, as it relates to complexity theory, is to analyze the complexity assumptions used as the basis for various cryptographic protocols and schemes. A central question is determining how to generate intractible instances of these problems upon which to implement an actual cryptographic scheme. The candidates for these instances must be platforms in which the hardness assumption is still reasonable. Determining if these group-based cryptographic schemes are quantum-safe begins with determining the groups in which these hardness assumptions are invalid in the quantum setting.

In what follows we address the quantum complexity of the {\it Hidden Subgroup Problem (HSP)} to determine the groups in which the hardness assumption still stands. {\it The Hidden Subgroup Problem (HSP) asks the following: given a description of a group $G$ and a function $f:G \rightarrow X$ for some finite set $X$ is guaranteed to be strictly $H$-periodic, i.e. constant and distinct on left (resp. right) cosets of a subgroup $H \leq G$, find a generating set for $H$.} 
It is important to note that Simon's problem of computing a XOR-mask, Shor's algorithm for factoring and finding the discrete log, Boneh's algorithm for finding a hidden linear function, and Kitaev's algorithm for the abelian stabilizer problem are all special cases of HSP. Therefore, the HSP is directly related to problems such as breaking one-time pad, discrete logarithm problem, graph isomorphism problem (which is now known to be in quali-polynomial), lattice-based problems, the problem for factoring for RSA. 

The classical complexity of HSP is known \cite{childs2017lecture}: {\it Suppose that $G$ has a set $\mathcal{H}$ of $N$ subgroups, such that $H_1 \cap H_2 \cap \ldots \cap H_{\mathcal{H}} = e_G$. Then a classical computer must make $\Omega(\sqrt{N})$ queries to solve the HSP.} The classical cases in which HSP is easy are the cases in which $G$ has only a polynomial number of subgroups, allowing brute-force for the function $f$ on all subgroups.

We provide a survey of results regarding the complexity of quantum algorithms for solving HSP in various group platforms. 
We also provide information on the relationship between HSP and other computational problems. These results provide insight into potential platforms for quantum safe cryptography, when the underlying hard problem is reducible to HSP.

\section{Group-based Cryptography} \label{subsec:gt}

Group-based cryptography could be shown to be post-quantum if the underlying security problem is NP-complete or unsolvable; firstly, we need to analyze the problem's equivalence to HSP, then analyze the applicability of Grover's search problem. Cryptanalysis based on a reduction to solving HSP creates some obstacles as the groups under consideration below are mostly infinite and do not have an efficient algorithm for HSP. In the following cryptosystems a connection to HSP can assist in the analysis of security.

For example in~\cite{hart2018practical} a practical cryptanalysis of WalnutDSA was proposed, a post-quantum cryptosystem using the conjugacy search problem (CSP) over braid groups that was submitted to the NIST competition in 2017 ~\cite{anshel2017walnutdsa}. It has been argued since the braid group does not contain any non-trivial finite subgroups, there does not seem to be any viable way to connect CSP with HSP. It has been shown there is no reduction connection between the CSP and HSP, \cite{WW10, WWCN2010}.  As for analysis via Grover's algorithm \cite{grover1996fast}, it has been mentioned that a majority of the time for signature verification in WalnutDSA is repeated E-Multiplications.

There are alternative group-theoretic problems and classes of groups which have been proposed for post-quantum cryptography. For example, the first proto-cryptosystem based on groups was proposed by Wagner-Magyarik in \cite{wagner1984public} for which the word choice problem was hard. Later on Flores-Kahrobaei-Koberda proposed right-angled Artin groups for various other cryptographic protocols \cite{flores2016cryptography}, \cite{flores2018algorithmic}. Eick and Kahrobaei proposed Polycyclic groups, using the Conjugacy Search Problem \cite{eick2004polycyclic} for cryptography. Later on Gryak-Kahrobaei wrote a survey and proposed other group-theoretic problems for consideration using polycyclic groups \cite{gryak2016status}. Kahrobaei-Koupparis \cite{kahrobaei2012non} proposed a post-quantum digital signature using polycyclic groups. Kahrobaei-Khan proposed a public-key cryptosystem using polycyclic groups \cite{kahrobaei2006nis05}. Habeeb-Kahrobaei-Koupparis-Shpilrain proposed the use of a semigroup of matrices with a semidirect product structure \cite{habeeb2013public}.

Thompson groups have been considered by Shpilrain-Ushakov based on the Decomposition Search Problem \cite{shpilrain2005thompson}. Hyperbolic groups have been proposed by Chatterji-Kahrobaei-Lu using properties of subgroup distortion and the Geodesic Length Problem \cite{chatterji2016cryptosystems}.  Free metabelian groups have been proposed based on the Subgroup Membership Search Problem by Shpilrain-Zapata \cite{shpilrain2006combinatorial}. Kahrobaei-Shpilirain proposed Free nilpotent p-groups for a semidirect product public key cryptosystem \cite{kahrobaei2016using}. Linear groups were proposed by Baumslag-Fine-Xu \cite{baumslag2006cryptosystems}. Grigorchuk groups, have been proposed by \cite{petrides2003cryptanalysis}. Groups of matrices, for a Homomorphic Encryption scheme were proposed by Grigoriev-Ponomarenko \cite{grigoriev2003homomorphic}.





\section{Relation of HSP to Other Computational Problems}
\label{sec:otherproblems}

Many computational problems are special cases of the HSP; an efficient algorithm for HSP over a certain group implies an efficient algorithm for some other computational problem. 
It is important to note that one method of determining an efficient quantum solution to a hard problem consists of reducing the problem to an instance of HSP over a group with a known efficient solution. This consists of determining the appropriate group $G$, the subgroup $H$ and the strongly $H$-periodic function $f$. For example, Simon's problem can be viewed as an instantiation of HSP over $G=\mathbb{Z}_2^n$ with a subgroup $H$ of order $2$. Duetsch's algorithm solves a variant of HSP where $H$ is either $\{0\}$ (f is balanced) or $\mathbb{Z}_2$ (f is constant). Shor's algorithm solves period finding as a special case of HSP, allowing for an efficient quantum algorithm for factoring and discrete log.
  
The graph automorphism (resp. isomorphism) problem can also be framed as an instance of HSP. To solve graph automorphism we consider HSP in the symmetric group on $n$ letters, $G = S_n$, any function $f$ which hides the trivial subgroup is an automorphism. Analogously the graph isomorphism problem is an instance of HSP over the wreath product $G = S_n \wr S_2$ \cite{kobler2012graph}. Also, solutions to HSP can solve the abelian stabilizer problem; when $G$ is acting on a finite set $X$ and where $\text{St}_G(x)$ is the stabilizer of $x$ we have that $f_x:G \rightarrow X$ can be defined such that $g \mapsto g(x)$ is strongly $\text{St}_G(x)$-periodic.

A solution to a particular instance of HSP is a solution to the hidden linear functions \cite{boneh1995quantum}; if $g$ is a permutation of $\mathbb{Z}_N$ and $h:\mathbb{Z} \times \mathbb{Z} \rightarrow \mathbb{Z}_N$ is such that $x,y \mapsto x + ay \mod N$, we have $f = g \circ h$ hiding $\langle (-a, 1)\rangle$. Additionally, self-shift-equivalent polynomials can be framed as an instance of HSP, in this case Grigoriev shows how to compute the hidden subgroup \cite{grigoriev1997testing}.

An efficient solution to HSP would imply an efficient solution to certain lattice problems. Specifically, the $g(n)$-Unique Shortest Vector Problem (USVP) is NP-hard for $g(n)=\text{O}(1)$, and has a polynomial time classical solution when $g(n)$ is large. The dihedral HSP, based on standard-method (found below) can be used to solve poly($n$)-USVP \cite{regev2004quantum}. HSP over the symmetric and dihedral groups are highly motivated open questions in post-quantum group-based cryptography.

Another, related, computational problem is the Hidden Shift Problem, which has been proposed as a basis for post-quantum cryptography in symmetric cryptosystems that are quantum-CPA secure \cite{alagic2017quantum}. Other than the use of a generalization of Simon's algorithm, and Kuperberg's algorithm discussed above, very little is known about the Hidden Shift Problem. Clearly this problem is closely related to HSP as some solutions coincide. It is important to note that constructions baed on this Hidden Shift problem have also remained quantum secure.

\section{Solution Methods} \label{sec:methods}

The standard method of solving HSP over $G$ performs the following steps. First, the algorithm queries the $H$-periodic function $f$ in superposition and discards the register which holds the output. This leaves the first register entangled in a hidden subgroup state, a superposition of coset representatives for some left traversal $K \subset G$. Following this, the state can be sampled using post-processing techniques to determine $H$. In the following we have $\vert gH \rangle = \vert H \vert^{-1/2} \sum_{h \in H} \vert gh \rangle$ as the coset state. This approach reduces the problem to a problem of quantum mechanics: how to distinguish the members of an ensemble of quantum hidden subgroup states.

$$\vert G \vert^{1/2} \sum_{g \in G} \vert g,0 \rangle \mapsto \vert G \vert^{1/2} \sum_{g \in G} \vert g,f(g) \rangle \mapsto \rho_H = \vert H \vert \vert G \vert^{-1} \sum_{g \in K} \vert gH \rangle \langle gH \vert$$

How do we measure the state? The problem of distinguishing these quantum states has some proposed solutions. Most namely, the often optimal solution entitled Pretty Good Measurement (PGM) can be used. An obstacle to performing PGM is the lack of an efficient QTF/CFT in the underlying group. For these instances we know of no efficient quantum algorithm for solving HSP.

\section{Results}

\paragraph{Finite Abelian and Finite Near-Abelian.} 

The infamous quantum algorithms of Simon and Shor provide quantum solutions to HSP in the abelian cases where $G=(\mathbb{Z}/2)^n$ and $G = \mathbb{Z}$ respectively. Shor's algorithm extends to the general abelian case as well, providing a polynomial time quantum algorithms with bounded error 
\cite{simon1997power, shor1999polynomial, kitaev1997quantum}. The probability of success can be improved to $1$ when $G$ is {\it abelian of smooth order}, i.e. if all prime factors of $\vert G \vert$ are at most $(\log \vert G \vert)^c$ \cite{brassard1997exact}. 

In the case that $G$ is nearly abelian, i.e. if the value $\kappa(G) = \{\cap_{H \leq G} N(H)\}$ where $N(H)$ is the normalizer for $H$ is sufficiently large there are established computational bounds on HSP. The size of this intersection relative to the group is a measure of the abelianness of $G$. 
Gavinsky \cite{gavinsky2004quantum} gave results to show that an efficient algorithm exists when $[G : \kappa(G)] = \text{poly}(\log \vert G \vert)$.

The HSP can be solved in polynomial time by a quantum algorithm to find hidden normal subgroups of solvable groups and permutation groups, finding hidden subgroups of groups with small commutator subgroup and of groups admitting an elementary Abelian normal 2-subgroup of small index or with cyclic factor group \cite{ivanyos2003efficient}.  Subexponential algorithms for HSP in any solvable group have been given by Friedl et al. \cite{friedl2014hidden}.

When $G$ is a known finite abelian group with a subgroup $H \leq G$, given black-box access to the $H$-hiding function $f$, we know that a quantum computer can uniquely and completely determine $f$ in $\text{poly}\log(\vert G \vert)$ time and query complexity. When $G$ is nearly abelian, or built from abelian parts, one can leverage this fact to obtain an efficient algorithm for HSP.

\paragraph{Finite, Non-Abelian.}

The finite non-abelian case of HSP is more elusive. 
Shor's algorithm extends to any group $G$ when $H$ is normal if quantum fourier transform (QFT) can be efficiently computed over the group \cite{hallgren2000normal}. The algorithm also extends to when $H$ has few conjugates, requiring the quantum character transform (QCT) over the group algebra $\mathbb{C}[G]$  \cite{grigni2001quantum}. This variation is not applicable when $H$ has many conjugates, as in some of the following cases. Alternatively, when $H$ is normal in $G$, a black-box group, generators for $H$ can be found in time polynomial in the input size + $v(G)$ \cite{ivanyos2003efficient} without requiring an efficient QFT over $G$.
Additionally, the quantum computation of the discrete log in semi-groups \cite{childs2014quantum} is an instance of HSP.

When the group $G$ is the (discrete) Heisenberg group 
it is sufficient to be able to distinguish cyclic subgroups of order $p$, $H_{a,b}$
. Thus, finding an arbitrary $H$ reduces to determining two parameters $a,b$, given the coset state produced by the standard method using $f$ which hides $H_{a,b}$. This can be efficiently computed with an overall success probability close to $\frac{1}{2}$.


In a more difficult case we consider instances of HSP in the dihedral group of order $2N$, $D_N$ 
where the function $f$ hides $H$ of order $2$. 
In this case $H$, a hidden reflection, has many conjugates in $G$ and the QCT based solution is not applicable. Kuperberg stated that {\it finding an arbitrary hidden subgroup $H$ of $D_N$ reduces to finding the slope of a hidden reflection} and provides a quantum algorithm with both time and query complexity of $2^{\text{O}(\sqrt{\log N})}$, applicable to $D_N$ for all values of $N$ but achieving an even tighter complexity bound for specific smooth values of $N$  \cite{kuperberg2005subexponential} . Kuperberg's algorithm also provides a solution to the hidden shift problem in an arbitrary finitely generated abelian group $G$. 
Regev improved upon the bounds of Kuperberg's original algorithm providing a polynomial space variation to the original superpolynomial space algorithm, which still achieves subexponential complexity \cite{regev2004subexponential}. 
Regev showed that an efficient solution to the dihedral HSP implies a quantum solution to lattice problems \cite{regev2004quantum}. 

When $G$ is a type of wreath product 
$W_n =  \mathbb{Z}_2^n \wr  \mathbb{Z}_2$, Roetteler et al. \cite{roetteler1998polynomial} provide a positive result for finding an efficient solution to the non-abelian HSP within $W_n$. This result is due to the existence of an efficient non-abelian QFT in $W_n$. Wreath product groups are in turn a subset of semi-direct product groups. When $G$ is (one of some groups that are) a  semidirect product of abelian groups, alternative efficient algorithms have been proposed. The polycyclic HSP has been addressed for $\mathbb{Z}_{p^k} \rtimes \mathbb{Z}_2$ for fixed prime power $p^k$ \cite{friedl2003hidden},  $\mathbb{Z}_q  \ltimes \mathbb{Z}_p$ with $q | (p-1)$ and $q = \frac{p}{\text{polylog}(p)}$, certain affine groups \cite{moore2004power}, $\mathbb{Z}_{p^r}^m \rtimes \mathbb{Z}_p$ \cite{inui2004efficient}, with $p \in \mathbb{P}$, $\mathbb{Z}_{p^r} \rtimes \mathbb{Z}_{q^s}$ where $p^r/q = \text{poly}(\log p^2$ where $p,q \in \mathbb{P}$ and $r,s \in \mathbb{N}$ \cite{gonccalves2011solution}, and $\mathbb{Z}_N \rtimes \mathbb{Z}_{}q^s$ where $N$ has a special prime factorization \cite{gonccalves2017efficient}. 


In general, when $G$ is a group of finite order the HSP problem has quantum query complexity of $\text{poly} (\log \vert G \vert)$ as shown by Ettinger, H{\o}yer and Knill \cite{ettinger2004quantum}; for any group $G$, $O(\log \vert G \vert)$ queries provides sufficient statistical information to solve HSP. This result provides no guarantees on computational complexity. The real problem is determining how to implement the queries efficiently, as well as how to control the amount of postprocessing required by the algorithm. In the case of the dihedral group, an algorithm with the lower bound on query complexity has been constructed, but the postprocessing required is exponential. In many cases the inefficiency of a proposed quantum algorithm is primarily due to the inefficiency of the required quantum measurement or post-processing within the group. 

\paragraph{Infinite.}
What seems to be an obstacle for infinite groups is that the quantum computer should assume the following state:
$\vert G \rangle = \vert G \vert ^{-1/2}\sum_{g \in G} \vert g \rangle.$
The meaning of this is clear for finite groups. 
The abelian infinite HSP was clearly first considered with Shor's algorithm, over $\mathbb{Z}$. In \cite{GKissinger}, infinite-dimensional HSP has been mentioned, particularly for infinite abelian groups $\mathbb{Z}^N$. Additionally, HSP has been defined and considered for infinite abelian groups of the form $\mathbb{R}^k \times \mathbb{Z}^l \times (\mathbb{R}/\mathbb{Z})^s \times H$ for some finite group $H$ \cite{eisentrager2014quantum}.  Other than the cases of $\mathbb{Z}^N$, $\mathbb{R^N}$, $\mathbb{T}^N$, and combinations of these in which an efficient algorithm exists, the infinite and continuous HSP has not been addressed within the literature for the {\it non-abelian} case.













\bibliographystyle{unsrt}
\bibliography{HSP-KH-DK-2.bib}

\end{document}